\documentclass[twocolumn,amssymb,amsmath,aps,prb,showpacs]{revtex4-1}
\usepackage{array}
\usepackage{graphicx}
\usepackage{float}

\begin{document}
\title{Competing magnetic interactions and magnetocaloric effect in Ho$_5$Sn$_3$}
\author{Suman Mondal$^1$}
\author{Pushpendra Yadav$^2$}
\author{Anan Bari Sarkar$^2$}
\author{Prabir Dutta$^1$}
\author{Saurav Giri$^1$}
\author{Amit Agarwal$^2$}
\email{amitag@iitk.ac.in}
\author{Subham Majumdar$^1$}
\email{sspsm2@iacs.res.in}
\affiliation{$^1$School of Physical Sciences, Indian Association for the Cultivation of Science, 2A \& B Raja S. C. Mullick Road, Jadavpur, Kolkata 700 032, India}
\affiliation{$^2$Department of Physics, Indian Institute of Technology Kanpur, Kanpur-208016, India}

\begin{abstract}
The rare-earth intermetallic compound Ho$_5$Sn$_3$ demonstrates fascinating magnetic properties which include temperature driven multiple magnetic transitions and  field driven metamagnetism. We address the  magnetic character of this exciting compound through a combined experimental and theoretical studies.  Ho$_5$Sn$_3$ orders antiferromagnetically below  $~28$ K, and shows further spin reorientation transitions at 15 K and 12 K.  We observe a sizable amount of low-temperature magnetocaloric effect in Ho$_5$Sn$_3$ with a maximum value of entropy change $\Delta S$ = -9.5 JKg$^{-1}$K$^{-1}$ for an applied field of $H$ = 50 kOe at around 30 K. The field hysteresis is almost zero above 15 K where magneto-caloric effect is important. Interestingly, $\Delta S$ is found to change its sign from positive to negative as the temperature is increased above about 8 K, which can be linked to the multiple spin reorientation transitions.  The signature of the metamagnetism is  visible in the $\Delta S$ versus $H$ plot. The magnetic ground-state, obtained from the  density functional theory based calculation, is susceptible to the effective Coulomb interaction ($U_{\rm eff}$) between electrons. Depending upon the value of $U_{\rm eff}$, the ground-state can be ferromagnetic or antiferromagnetic. The compound shows large relaxation (14\% change in magnetization in 60 min) in the field cooled state with a logarithmic time variation, which may be connected to the competing magnetic ground-states observed in our theoretical calculations. The competing magnetic ground-states is equally evident from the small value of the paramagnetic Curie-Weiss temperature.
\end{abstract}
\maketitle

%%%%%%%%%%%%%%%%%%%%%%%%%%%%%%%%%%%%%%%%%%%%%%%%%%%%%%%%%%%%%%%%%%%%%%%%%%%%%%%%%%%%%%%%%%%%%%%%%%%%%%%%%%%%%%%%%%%%%%%%%%%%%%%%%%%%%%%%%%%%
\section{Introduction}
\label{CS}
Rare-earth (RE) based intermetallic compounds continue to be at the forefront of active research due to their intriguing magnetic and electronic properties, which concern both fundamental and applied aspects of magnetism~\cite{zhang,buschow1,zhao}. In recent times, RE based compounds have gained immense importance due to their applications in the field of magnetic refrigeration, magnetic sensors, memory devices as well as for the development of permanent magnets~\cite{gschneidner1,Li2020,COEY2020119,mram}. All these functional properties of RE-intermetallics are closely linked to the large localized moment of the RE atom as well as the excellent metallic character with a large density of states at the Fermi level. The latter enables strong magnetic interaction through indirect Ruderman-Kittel-Kasuya-Yosida (RKKY) interaction mediated by the free electrons. 
\par
 A binary or a ternary RE-intermetallic (REI) may contain a transition metal element, a post-transition $sp$ element (Z) or both. Depending upon  transition metal or Z and the stoichiometry, the REIs can form in diverse crystallographic structures. An essential aspect of REIs is magneto-crystalline anisotropy, which has its origin in the relatively strong spin-orbit coupling (SOC) in the system~\cite{anisotropy}. The anisotropy varies in REIs depending upon the lattice symmetry and the SOC. 

\par
The observation of giant magnetocaloric effect (MCE) in Gd$_5$Ge$_2$Si$_2$~\cite{pecharsky} has opened up a new avenue for research in the REIs~\cite{li}. MCE is a magneto-thermal phenomenon in which the temperature of a sample is changed when the sample is exposed to a variable external magnetic field. Isothermal magnetic entropy change ($\Delta S$) and the corresponding adiabatic temperature change ($\Delta T$) are the two experimentally obtained parameters that measure the MCE of a sample. The magnetic refrigeration~\cite{tishin1,gschneidner1,pecharsky1,pecharsky2,provenzano1,provenzano2,provenzano3} based on MCE is environment-friendly, safer, compact and devoid of mechanical motion. Metamagnetic transition plays a vital role towards the observation of giant MCE in materials such as Gd$_5$Ge$_2$Si$_2$. 
\par
Among the binary and pseudo-binary heavy RE compounds and alloys, Gd$_5$Ge$_4$ based compositions with 5:4 stoichiometry are the most widely studied due to their exotic ground state, metastability, phase coexistence, magneto-structural instability and above all giant MCE~\cite{gd5ge4}. In contrast, the RE$_5$X$_3$ (X is a non-magnetic element) compositions are much less studied~\cite{Drzyzga,Tb5Si3,Ho5Si3,Er5Si3,gd5sb3,gd5sn3,Szade, Drzyzga,Er5Sn3,Tb5Sn3,R5Pb3,canepa,dy5in3}. This paper focuses on the magnetic and magnetocaloric properties of a Ho-based REI compound Ho$_5$Sn$_3$. Unlike Gd, Ho has non-zero orbital angular momentum ($L \neq$ 0), which can provide sizable SOC and associated  magneto-crystalline anisotropy in the system.  Furthermore, Ho$_5$Sn$_3$ has a hexagonal crystal structure  that is structurally anisotropic. Thus, we expect interesting magnetic behaviour in this composition originating from the mutual interplay of structure, orbital, spin and electronic degrees of freedom~\cite{re-intermetallics}.
\par
Earlier work on Ho$_5$Sn$_3$ indicates that it attains a long-range magnetic ordered state below $T_N$ = 22 K with an incommensurate antiferromagnetic (AFM) ground state~\cite{ho5sn3}. Two other spin reorientation transitions are observed at $T_1$ = 16 K and $T_2$ = 12 K. The magnetic structure between $T_N$ and $T_1$ is a sine modulated incommensurate one with a single propagation vector ${\bf k_1}$ = (0,0,0.6893). Below $T_1$, the magnetic structure consists of two propagation vectors, namely ${\bf k_1}$ = (0,0,0.6893) and an additional commensurate one ${\bf k_2}$ = (0,0,1). At temperatures below $T_2$, the structure consists of an additional incommensurate propagation vector ${\bf k_3}$ = (0.5956,0,0). The sample also shows a metamagnetic transition under the application of $\sim$ 20 kOe of field~\cite{ho5sn3}.
\par
To further explore the  magnetic properties of Ho$_5$Sn$_3$, we present here a joint experimental and theoretical investigation on the compound. The large moment at the Ho site and the presence of metamagnetism prompted us to study the MCE and its fundamental magnetic properties.  The paper is organized as follows. In Sec.~\ref{method}, we discuss the crystal structure and details of the experimental and computational techniques. In Sec.~\ref{MGS}, we discuss electronic structure and possible magnetic ground state from the first principle calculations. The experimental results for the magnetic properties are discussed in Sec.~\ref{MP}. In Sec.~\ref{MCE}, we focus on the magneto-caloric effect in Ho$_5$Sn$_3$. Finally, we summarized and discussed our findings of in Sec.~\ref{discussions}.
%%%%%%%%%%%%%%%%%%%%%%%%%%%%%%%%%%%%%%%%%%%%%%%%%%%%%%%%%%%%%%%%%%%%%%%%
\begin{figure}[t]
\centering
\includegraphics[width =0.9\linewidth]{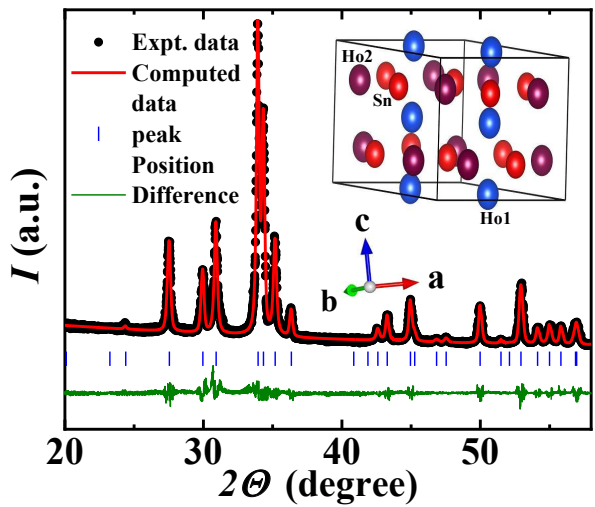}
\caption {XRD pattern at room temperature with Rietveld refinement curve (solid line). Inset: Perspective views of the crystal structure viewed along the $b$ axis.}
\label{xrd}
\end{figure}
%%%%%%%%%%%%%%%%%%%%%%%%%%%%%%%%%%%%%%%%%%%%%%%%%%%%%%%%%%%%%%%%%%%%%%%%%%%%
\section{Crystal structure and methodology}
\label{method}
Ho$_5$Sn$_3$ has a layered crystal structure with a hexagonal symmetry [space group $P6_3$/mcm (193)] and its lattice parameters are $a$ = 8.845~{\AA} and $c$ = 6.452~{\AA}~\cite{ho5sn3}. The compound has a total of 16 atoms in its unit cell of which 10 are Ho atoms and 6 are Sn atoms. Ho has two crystallographically inequivalent sites, namely 4$d$ (1/3, 2/3, 0) and 6$g$ ($\xi_{Ho}$, 0, 1/4).  Here $\xi_{Ho}$ is the $x$ coordinate of the 6$g$-Ho, and its value is close to 0.241. The Ho atom at the 4$d$ site has 14 neighbours, including two Ho(4$d$), six Ho(6$g$) and six Sn. On the other hand, Ho at the 6$g$ site has 15 neighbours, which comprise of four Ho(4$d$), six Ho(6$g$) and six Sn. There are multiple Ho-Ho bondings, such as Ho(4$d$)-Ho(4$d$), Ho(4$d$)-Ho(6$g$) and Ho(6$g$)-Ho(6$g$) with different bond lengths. Since the sign and strength of the RKKY interaction depend upon the distance between magnetic atoms, the different bond-lengths in Ho$_5$Sn$_3$ can give rise to very intriguing  magnetic interactions in Ho$_5$Sn$_3$. The competing magnetic interactions can  cause magnetic frustrations~\cite{Bandyopadhyay,Bandyopadhyay2}. 

\par
Polycrystalline sample of Ho$_5$Sn$_3$ were prepared by standard argon arc melting technique. The crystallographic structure was investigated by room temperature powder x-ray diffraction (XRD) experiment using Cu-K$_{\alpha}$ radiation. The obtained XRD pattern was analysed using the Rietveld refinement technique with MAUD program package as shown in Fig.~\ref{xrd}. All reflections  could  be  indexed  on  the  basis  of  the hexagonal system P6$_3$/mcm with lattice constants $a$ = 8.841~\AA~ and $c$ = 6.635~\AA. In our refinement, 6$g$- Ho is found to occupy (0.24(2),0,0.25), while 6$g$ Sn occupies (0.60(1),0,0.25) position. These values are consistent with the values reported previously~\cite{ho5sn3}.  The magnetization ($M$) of the samples was measured using a quantum design SQUID magnetometer (MPMS XL 7, Evercool model) in the temperature ($T$) range 2-300 K with maximum applied magnetic field ($H$) of 70 kOe. The resistivity ($\rho$) was measured using standard four probe technique by a constant current source and a nano-voltmeter attached to a closed cycle cryogenic system..
%%%%%%%%%%%%%%%%%%%%%%%%%%%%%%%%%%%%%%%%%%%%%%%%%%%%%%%%%%%%%%%%%%%%%%%%%%%%%
	\begin{figure*}[t!]
		\includegraphics[width =\linewidth]{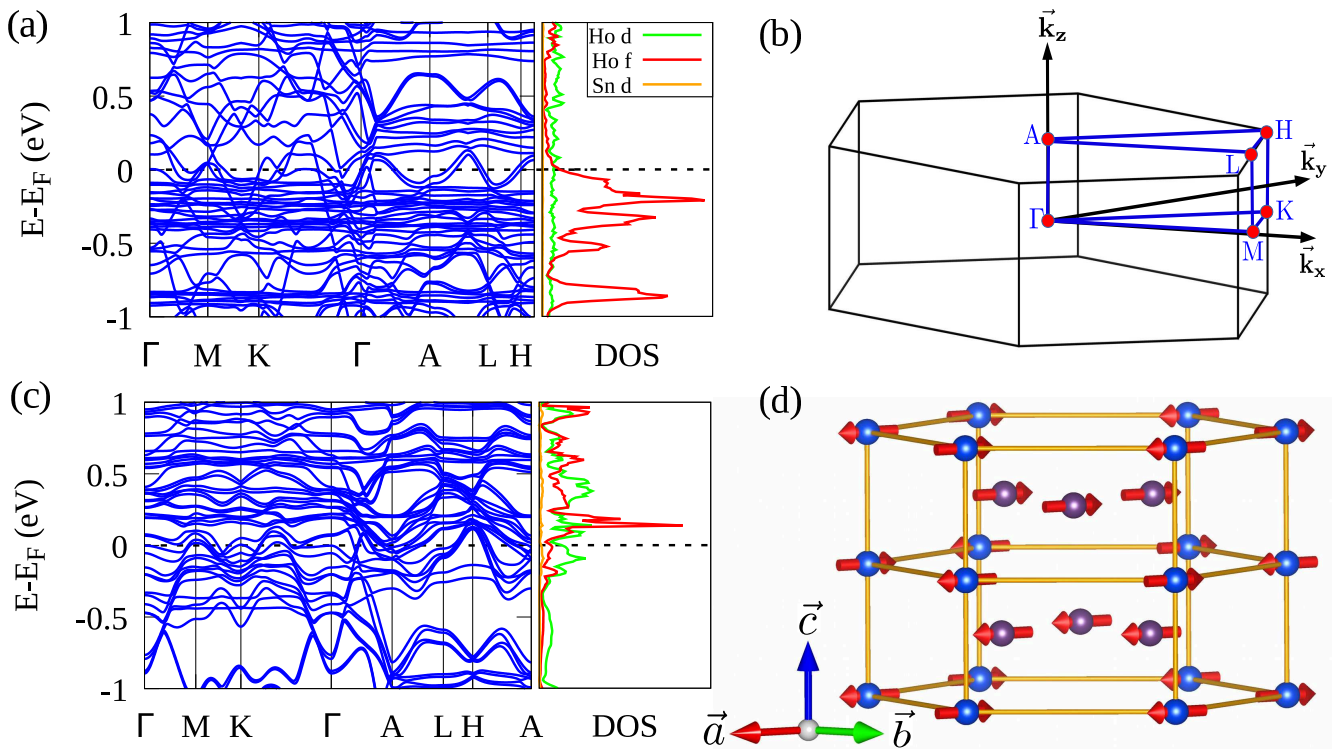} 
				\caption{(a) The non-magnetic electronic band structure and density of states of Ho$_5$Sn$_3$ with spin-orbit coupling. Most of the localized bands in vicinity of the Fermi surface arise from the Ho-$d$ orbitals and Ho-$f$ orbitals.  (b) The Brillouin zone with different high-symmetry points marked in red dots. (c) The electronic band structure and density of states (DOS) in the presence of SOC for the in-plane AFM configuration, which is the ground state for $U = 11$ eV. (d) The corresponding magnetization configuration associated with the Ho$_1$ (blue) and the   Ho$_2$ (purple) atoms.}
			\label{bz}
\end{figure*}
%%%%%%%%%%%%%%%%%%%%%%%%%%%%%%%%%%%%%%%%%%%%%%%%%%%%%%%%%%%%%%%%%%%%%%%
%%%%%%%%%%%%%%%%%%%%%%%%%%%%%%%%%%%%%%%%%%%%%%%%%%%%%%%%%%%%%%%%%%%%%
\begin{figure*}[t!]
\centering
\includegraphics[width = 12 cm]{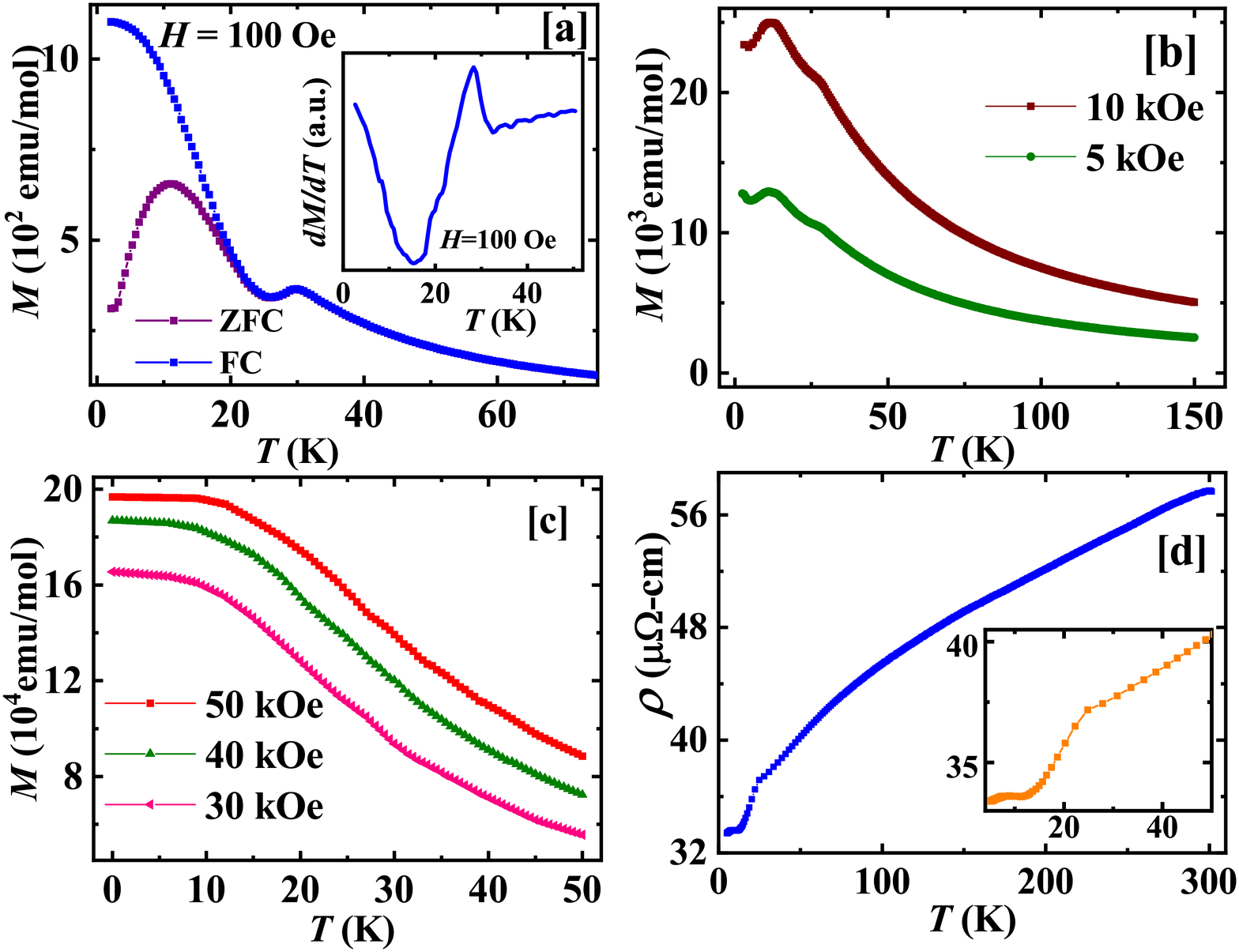}
\caption {(a) The magnetization as a function of the temperature, with zero field-cooled (ZFC) and field-cooled (FC) protocols, measured under $H$ = 100 Oe. The inset shows the temperature derivative of $M(T)$, with a clear minimum around  15 K and a maximum at 28 K. (b) shows the magnetization versus temperature data (ZFC) measured under 5 and 10 kOe. (c) depicts the magnetization versus temperature data (ZFC) measured under 30, 40 and 50 kOe.  (d) The resistivity as a function of temperature, with the inset showing an enlarged view of the low temperature anomalies.}
\label{mt}
\end{figure*}
%%%%%%%%%%%%%%%%%%%%%%%%%%%%%%%%%%%%%%%%%%%%%%%%%%%%%%%%%%%%%%%%%%%%%%%

\par
We performed first principle calculations based on density functional theory to explore the magnetic ground-state and electronic properties, as implemented in the Vienna ab-initio simulation package (VASP) ~\cite{PhysRevB.54.11169,PhysRevB.59.1758}. The exchange correlation effects were treated within the generalized gradient approximation (GGA)~\cite{PhysRevLett.77.3865}. Since GGA often fails to describe the localized electrons correctly~\cite{Torelli-2019,PhysRevB.102.035420}, we considered on-site Coulomb interaction for Sn-$d$ orbitals and Ho-$f$ orbitals. Within the GGA + $U$ formalism, we used $U$ = 3.5 eV for Sn atoms and $U$ = 11 eV for Ho atoms, respectively. We used 300 eV as the kinetic energy cut-off for the plane-wave basis set. To perform Brillouin zone (BZ) integrations, we used a $\Gamma$-centered $9\times 9 \times 9$ Monkhorst $k$ mesh~\cite{PhysRevB.13.5188}. A tolerance of $10^{-5} $ eV is used for electronic energy minimization. The magnetic calculations for the system were carried out  by assuming the Sn-atoms to be non-magnetic and the Ho-atoms to be magnetic. %We performed the magnetic calculations with and without the inclusion of SOC.

\begin{table*}[!]
	\caption{The calculated ground-state energy in Ho$_5$Sn$_3$ (per unit cell) for different magnetic configurations with all moments aligned ferromagnetically or anti-ferromagnetically (denoted as E$_{FM/AFM}$) along the different crystal axes ($a/c$). We find that the ground-state is sensitive to the chosen values of the Hubbard interaction parameter $U_{\rm eff}$.}

	\centering
	\begin{tabular}{>{\centering\arraybackslash}p{2.4cm}>{\centering\arraybackslash}p{2cm}>{\centering\arraybackslash}p{2cm}>{\centering\arraybackslash}p{3cm}>{\centering\arraybackslash}p{3cm}>{\centering\arraybackslash}p{2cm}}
		\hline \hline
		$U_{\rm eff}$ & Spin axis & E$_{FM}$(eV) & E$_{AFM}$(eV) & E$_{FM}$-E$_{AFM}$(eV) & Groung state \\
		\hline \hline
		11 & a &  -115.5458 &-116.7614& 1.1256 & AFM$^a$  \\
		
		 & c &  -116.3889 &-116.4257& 0.0368 & \\ 
		\hline
		
		10 & a &  -114.7937 & -113.8796& -0.9141 & FM$^a$  \\
		
		 & c &  -114.3608 &-114.3499& -0.0109 &  \\ 
		\hline
		
		9 & a &  -113.2535 &-113.1311& -0.1224& FM$^a$  \\
		
		 & c & -112.8051 & -112.9042 & 0.0991 &  \\ 
		\hline
		
		8 & a &  -111.8738 &-111.9824 & 0.1086& AFM$^a$  \\
		
		 & c & -111.6177 & -111.5375 &-0.0802 &  \\ 
		\hline	
		
		7 & a &  -111.6945 &-111.4358 & -0.2587& FM$^a$  \\
		
	      & c & -110.7748 & -110.7105&-0.0643 &  \\ 
		\hline
		
	\end{tabular}
	\label{T1}
\end{table*}

\section{Electronic structure and possible magnetic ground-state}
\label{MGS}
The non-magnetic bandstructure of Ho$_5$Sn$_3$ along different high-symmetry direction and its associated Brillouin zone is shown in Fig.~\ref{bz}(a)-(b). The non-magnetic bands and the corresponding density of states (DOS) in Fig.~\ref{bz}(a) clearly show its metallic characteristic. The presence of unpaired electrons in the Ho-$f$ orbitals indicates the strong possibility of Ho$_5$Sn$_3$ hosting a magnetic ground-state. Due to the presence of on-site Hubbard $U_{\rm eff}$ in Sn $d$ and Ho $f$ orbitals, the contribution from these orbitals in the DOS is shifted away from the Fermi energy (not shown), keeping only the Ho $d$ orbital contribution near the Fermi energy.

\par
To explore some simple possibilities for the magnetic ground-state, we systematically calculate the ground-state energies for considering different spin configurations in the ferromagnetic (FM) and AFM states with all the moments aligned along the crystallographic $a$ and $c$ axis. Even for such simple configurations, we find that the exact ground-state is very sensitive to the precise value of the short-range interaction parameter (the Hubbard $U_{\rm eff}$) used for the Ho-$f$ orbitals in the calculations. Interestingly, for several choices of the $U_{\rm eff}$ values, we find that different magnetic configurations (such as FM$_{a,c}$, AFM$_{a,c}$) are energetically very close  (see Table~\ref{T1} for details). For example, with $U_{\rm eff} = 11$ eV we found that the in-plane AFM state is the magnetic ground-state. Its magnetic configuration (AFM$_a$) and the resulting band structure are shown in panels (d) and (c) of Fig.~\ref{bz}. However, while the energy difference between FM and AFM configurations with moments aligned  along the $a$ axis is large, $\approx$ 1.12 eV, the difference in the energy configurations with moment along $c$ direction is $\approx$ 37 meV. This large difference in energy while aligning the moment along two different directions indicates the possibility of {\it non-collinear magnetism} or even other magnetic states such as ferri-magnetism in the system. Given this scenario, it is difficult to pin down the exact magnetic ground-state configuration from first-principle calculations. 
\par

%%%%%%%%%%%%%%%%%%%%%%%%%%%%%%%%%%%%%%%%%%%%%%%%%%%%%%%%%%%%%%%%%%%%%%%%%%%%%%%%%%%%%%%%%%%%%%%%%%%%%%%%%%%%%%%%%%%%%%%%%%%%%%%%%%
\section{Magnetic properties}
\label{MP}
In Fig.~\ref{mt}(a), we show the $M$ vs $T$ data in the presence of $H$ = 100 Oe, obtained using  the zero-field-cooled (ZFC) and the field-cooled (FC) protocols. Ho$_5$Sn$_3$ shows a clear peak at 28 K, indicating the onset of AFM order. This value of $T_N$ = 28 K is slightly higher than the value reported in the previous work~\cite{ho5sn3}. On cooling, the ZFC curve deviates from FC curve at  18 K, and the ZFC data show a peak at 12 K, which indicates a second spin reorientation transition occurring at $T_2$. The first derivative curve ($dM/dT$ vs $T$) of the FC data show a peak at $T_N$ = 28 K, and a minimum is observed at 15 K [see inset of fig. 3 (a)]. The minimum is likely to be associated with the first spin reorientation transition ($T_1$) found in the previous work~\cite{ho5sn3}. We have fitted the high-$T$  magnetic susceptibility ($\chi$ = $M/H$) with a Curie-Weiss law (not shown here). The paramagnetic moment and Curie-Weiss temperature, obtained from our $\chi(T)$ data measured at $H$ = 1 kOe, are found to be $\mu_{\rm eff} = 11.4~\mu_B$ and $\theta_p =9$ K, respectively. The positive value of  $\theta$ indicates the presence of a sizable  FM correlation  in the system. In a previous report, $\theta_p$  was reported to be  -1 K  by fitting  $\chi(T)$ data between 30 to 100 K. This value is slightly different from our result ($\theta_p =9 $ K). We obtained $\theta_p $  from high temperature region, and it is generally considered to be more genuine. The value of $\theta_p$ in our study  is much smaller than $T_N$, indicating the competition  between FM and AFM correlations. 

\par
If we compare the $M$ versus $T$ recorded at different values of $H$ [see Figs.~\ref{mt}(b) and (c)], the magnetic anomalies observed at 12 K and 28 K gradually turn weaker with increasing $H$. At 50 kOe, $M(T)$ data show only a change in slope (instead of peak) at these transition points. Interestingly, $M$ becomes almost $T$ independent below about 10 K in the 50 kOe data. These changes can be attributed to the metamagnetic transition observed in the sample around 25 kOe. We would like to mention that no thermal hysteresis is seen in the $M(T)$ data across the peaks around 28 and 12. This indicates that the transitions are primarily second order in nature.   
%%%%%%%%%%%%%%%%%%%%%%%%%%%%%%%%%%%%%%%%%%%%%%%%%%%%%%%%%%%%%%%%
\begin{figure}[t!]
\centering
\includegraphics[width = 8 cm]{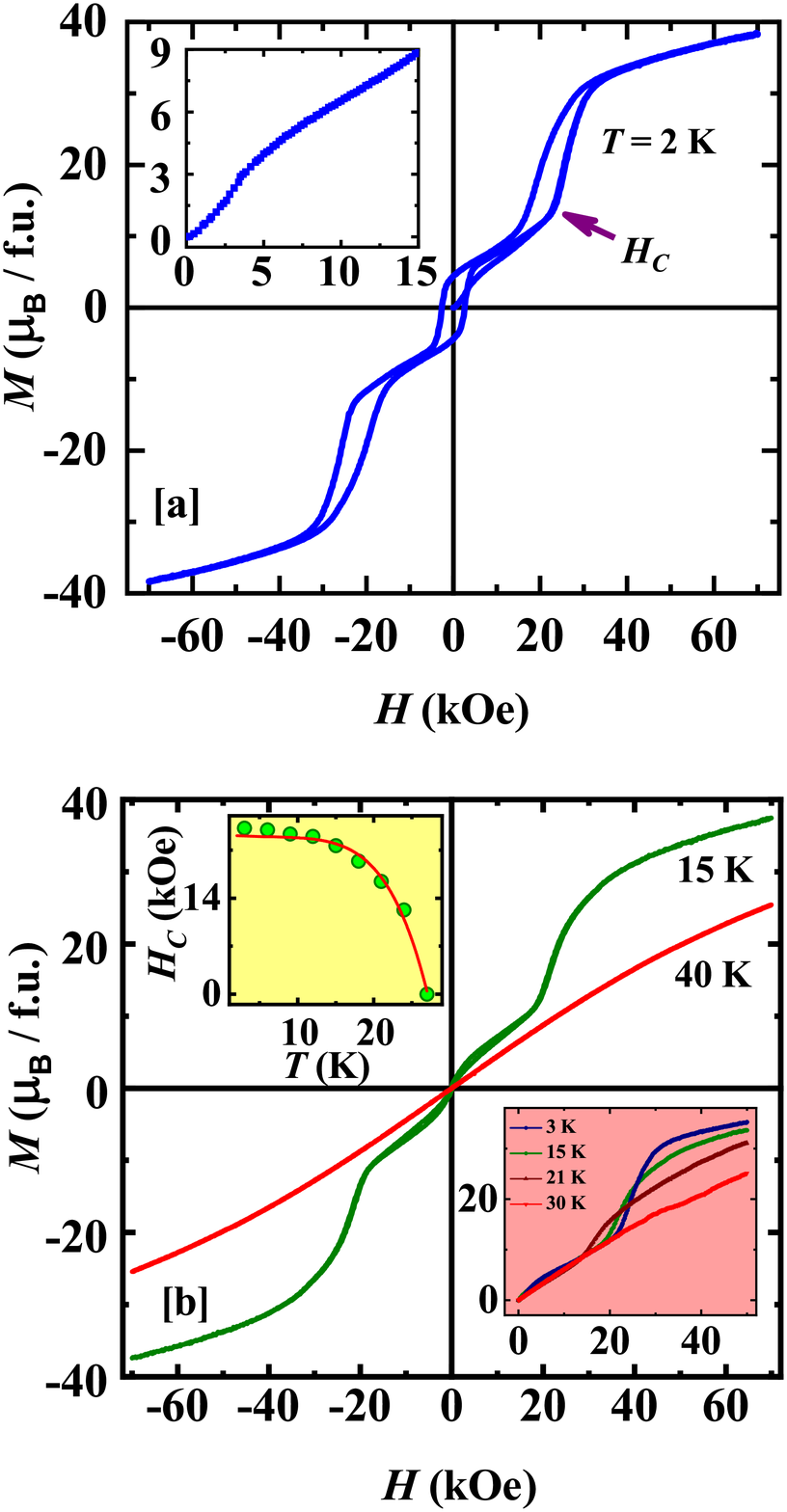}
\caption {(a) Magnetization versus  magnetic field data measured at $T =$ 2 K. The inset of (a) shows the low field data, clearly highlighting a change in the slope around 4 kOe. (b) shows the magnetization data measured at 15 K and 40 K respectively.  Lower inset of (b) depicts $M$-$H$ curves measured at various temperatures. The upper inset of (b) shows the $T$ variation of the critical field of metamagnetic transition.}
\label{mh}
\end{figure}
%%%%%%%%%%%%%%%%%%%%%%%%%%%%%%%%%%%%%%%%%%%%%%%%%%%%%%%%%%%%%%%%%
\par
The  $\rho$ versus $T$ plot, shown in Fig.~\ref{mt}(d), indicates clear anomalies around 25 K and a weak peak around 6 K. The 25 K anomaly is likely to be related  to the long-range magnetic order occurring at $T_N$. $\rho(T)$ is found to decrease with the decrease of $T$ which indicates the metallic nature of the sample. It further supports our theoretical calculation of the metallic state.

%%%%%%%%%%%%%%%%%%%%%%%%%%%%%%%%%%%%%%%%%%%%%%%%%%%%%%%%%%%%%%%%%%%%%%%%%%%%%%%%%%%%%%%%%%%%%%%%%%%%%%%%%%%%%%%%%%%%%%%%%%%%%%%%%%
\par
Figure~\ref{mh}(a) shows the isothermal $M$  versus $H$ data recorded at 2 K.  Clearly $M(H)$ has  a step-like anomaly around 25 kOe (marked by $H_C$), which can be assigned to the metamagnetic transition. In our experiments, the feature of the metamagnetism  is much more prominent and sharp as than the previous report~\cite{ho5sn3}. Two separate field hysteresis regions are observed in the $M(H)$ data, namely between $H$ = $\pm$ 10 kOe (around the origin) and also between 10 to 30 kOe (similarly, between -10 to -30 kOe in the 3rd quadrant). The magnetization curve does not saturate even at the highest applied field of 70 kOe, and the value of $M$ at the highest $H$ is found to be 39 $\mu_B$/f.u. This indicates that the system does not attain a completely spin aligned state along the field direction. 
\par
Interestingly, the nature of the $M(H)$ curve is found to be markedly different at 15 K, which is close to $T_1$ of the sample.  At 15 K, the hysteresis is weak, and the coercive field is almost zero [Fig.~\ref{mh}(b)]. The step-like metamagnetic transition is still present but at a lower value of $H$.  At 40 K, which is above $T_N$, the $M(H)$ curve is devoid of any hysteresis [Fig.~\ref{mh}(b)], but shows some non-linearity. In the lower inset of Fig.~\ref{mh}(b), we have depicted several isothermal magnetization curves (in the first quadrant) recorded at different constant temperatures. It is clear that the metamagnetic field decreases with increasing $T$, and its variation with $T$ is depicted in the upper inset of Fig.~\ref{mh}(b). The $H_C$ versus $T$ often follows an empirical  relation, $H_C = H_{0}[1-(T/T_0)^n]$, where $n$ is an exponent, while $T_0$ is the ordering temperature, and $H_{0}$ is the critical field at $T$ = 0.  The best fit between  (see the solid line in the inset) gives us the value of $n$ to be 5.2, $T_0$ = 25 K and $H_{0}$ = 23 kOe.

%%%%%%%%%%%%%%%%%%%%%%%%%%%%%%%%%%%%%%%%%%%%%%%%%%%%%%%%%%%%%%%%%%%5
\begin{figure}[t!]
\centering
\includegraphics[width = 8 cm]{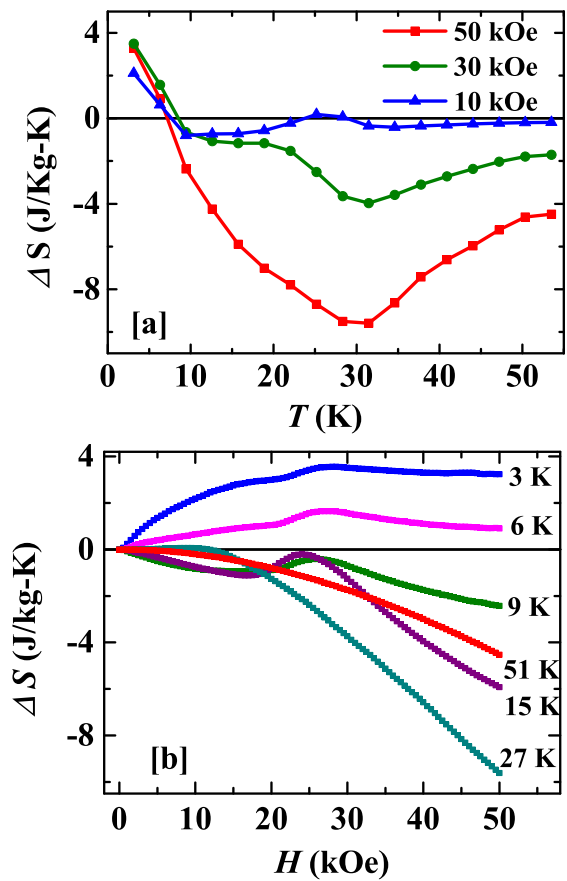}
\caption {(a) shows temperature variation of magnetocaloric effect ($\Delta S$) for different magnetic fields. (b) depicts isothermal variation of $\Delta S$ with $H$  for different temperatures.}
\label{mce}
\end{figure}
%%%%%%%%%%%%%%%%%%%%%%%%%%%%%%%%%%%%%%%%%%%%%%%%%%%%%%%%%%%%%%%%%%%%%%%%%%%%%%5
If we look at the low-temperature $M(H)$ data [Fig.~\ref{mh} (a)], the magnetization suddenly  rises at the onset point of metamagnetsim, $H_C$. On further increase of $H$, $M$  gently increases. This possibly indicates a spin-flop type transition. In a spin flop transition for a collinear AFM system with relatively weak magneto-crystalline anisotropy, the sublattice magnetization suddenly attains a direction perpendicular to the easy axis of magnetization at $H_C$. With increasing field beyond $H_C$, magnetization gradually tends towards the direction of the field ~\cite{nio,bogdanov,gignoux}. However,  Ho$_5$Sn$_3$ have complex sine modulated structure with incommensurate magnetic propagation vector, and such a simplified picture of spin-flop transition may not hold good. It is to be noted that above $H_C$, all the peaks in the $M(T)$ data smooth out. Since we do not see saturation in the $M(H)$ data even at 70 kOe, a completely aligned high field state can be ruled out. We also observe a change in slope in the $M(H)$ data around 4 kOe [see inset of Fig.~4(a)], which is a signature of additional low field metamagnetism.

%%%%%%%%%%%%%%%%%%%%%%%%%%%%%%%%%%%%%%%%%%%%%%%%%%%%%%%%%%%%%%%%%%%%%%%%%%%%%%%%%%%%%%%%%%%%%%%%%%%%%%%%%%%%%%%%%%%%%%%%%%%%%%%%%%%%%%%%%%%%%%% 
\section{Magnetocaloric Effect}
\label{MCE}
The presence of metamagnetic transition prompts us to calculate MCE around the magnetic transitions. Around a metamagnetic transition, $H$ can induce a large change in the spin entropy due to the change in spin structure. We measured MCE in terms of the change in entropy ($\Delta S$) due to the variation of $H$. We recorded a series of isotherms around the magnetic transition and used the Maxwell's relation,
\begin{equation}
\centering
 {\Delta}S(0{\rightarrow}H_0)=\int^{H_0}_0\left(\frac{{\partial}M}{{\partial}T}\right)_HdH~. 
 \label{deltas}
\end{equation}
Here, ${\Delta}S(0{\rightarrow}H_0)$ denotes the change in entropy when the magnetic field is changed from 0 to $H_0$.  Figure~\ref{mce}(a) shows the variation of $\Delta S$ as a function of $T$ with different values of $H_0$. The maximum value of MCE is -9.5 Jkg$^{-1}$K$^{-1}$ with a peak around $T_N$. Relative cooling power (RCP) is an important parameter which can be used to judge the potency of a particular material to be used for magnetic refrigeration. If $\Delta  S_{\rm max}$ is the peak value of the MCE in the $\Delta S$ versus $T$ plot, RCP = $|\Delta  S_{\rm max}|\times\delta T_{\rm fwhm}$. Here $\delta T_{\rm fwhm}$ is the full width at half maxima of the peak in the $\Delta S$ versus $T$ plot. The calculated  RCP of Ho$_5$Sn$_3$ is  344  Jkg$^{-1}$, and this value  is relatively high among the REIs. We have compared the values of $\Delta  S_{\rm max}$ and RCP of certain potential magnetocaloric  materials among REIs with the peak position of $\Delta S$ lying between 6-28 K (see table~\ref{mce-comp}).  Ho$_5$Sn$_3$ with similar MCE and RCP values belongs to the same league of materials.
\par
In the case of Ho$_5$Sn$_3$, $\Delta S$ is positive (inverse MCE) for $T <$ 8 K, and it turns negative above this temperature. The magnitude of $\Delta S$ at 9 K  and above increases with increasing $H$, although there are some complex $H$ dependence around 25 kOe due to the metamagnetic transition. The inverse MCE [see 3 K and 6 K data in fig. 5 (b)] increases monotonically with $H$ till about 30 kOe. Above this field $\Delta S$ tend to decrease weakly with $H$. The positive  MCE is not unusual in an AFM or ferrimagnetic state~\cite{ranke}, because an applied  field can  introduce more spin disorder leading to the increase in the value of magnetic entropy.  The material like Ho$_5$Sn$_3$, where mixed magnetic interactions are present~\cite{wjhu}, is more liable to show such field-induced disorder. 

\begin{table*}
\begin{tabular}{|c|c|c|c|c|}
                    \hline \hline 
Sample  & $\Delta S_{max}$ &$T_{max}$& RCP & [Ref] \\

        & Jkg$^{-1}$K$^{-1}$ & K & Jkg$^{-1}$&\\
          \hline 
					\hline														
		Ho$_5$Pd$_2$&-18 & 28 & 877&~\onlinecite{lingwei}  \\ 
		GdCo$_2$B$_2$ &-17.1 & 25 & 462& ~\onlinecite{lingwei}   \\ 																
    Ho$_2$Ni$_{0.95}$Si$_{2.95}$ &-23.25 & 10 & 460&~\onlinecite{pakhira2}  \\ 
		HoCuSi &-33.1 & 7& 385&~\onlinecite{chen}  \\ 
		ErFeSi &-14.2 &22 & 365&~\onlinecite{zhang3}  \\ 	
		ErMn$_2$Si$_2$&-25.2 & 5& 365 &~\onlinecite{lingwei2}   \\ 
		GdPd$_2$Si&-15 &17 & 320 &~\onlinecite{lingwei}   \\ 
		ErRu2Si$_2$ &-17.6 &6 & 262 &~\onlinecite{lingwei3}  \\ 
		PrCo2B$_2$ &-8.1 & 16 & 104& ~\onlinecite{lingwei} \\
		Ho$_5$Sn$_3$ &-9.5&28&344&this work\\		
\hline 
 \end{tabular}
\caption{Values of maximum entropy change ($\Delta S_{max}$), temperature where peak value of entropy change is observed ($T_{max}$) and the relative cooling power (RCP) for various low -temperature magnetocaloric materials. Here all the data for the magnetic field change from zero to 50 kOe.}
\label{mce-comp}
\end{table*} 

\par
To address the  magnetic states of the system through our MCE measurements, we have plotted the $H$ variation of $\Delta S$ as shown in Fig.~\ref{mce}(b). The change in sign of MCE curve with temperature is also clear from the field variation of $\Delta S$. We find a peak around $H_C$ which shifts towards lower field values with increasing $T$. The 3 and 6 K plots are found to be positive at all fields, with a peak around 28 kOe. The 9 K and 15 K plots show negative values of $\Delta S$ with a peak centred around 28 kOe. For $T>T_N$, the variation of $\Delta S$ with $H$ is monotonous and we find that it varies as $H^2$ (plot for the fitting is not shown here) as expected for a paramagnetic state. 

\par
\begin{figure}[t!]
\centering
\includegraphics[width = 8 cm]{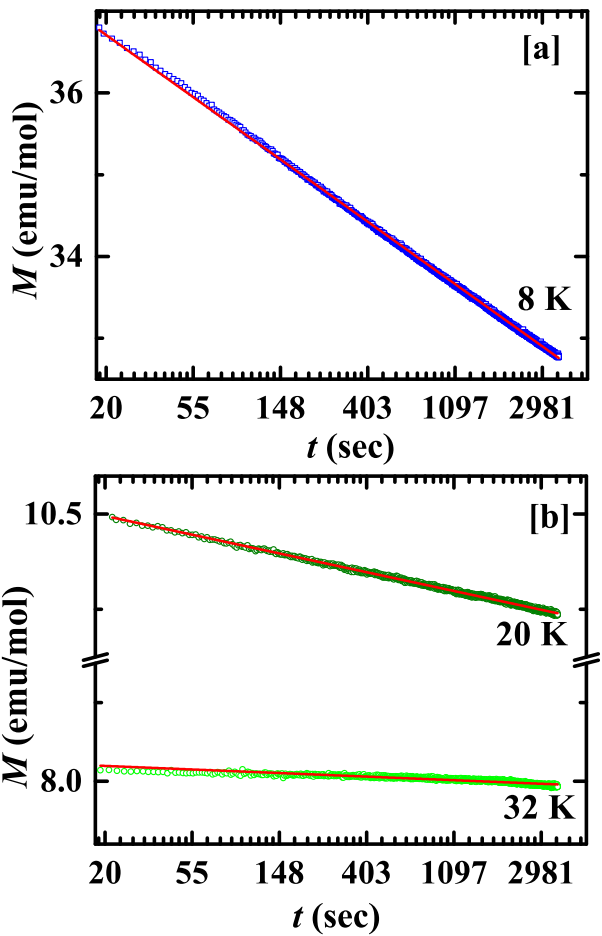}
\caption { (a) and (b) show time variation of magnetization at different temperatures.}
\label{relx}
\end{figure}

Considering complex magnetic interaction pathways in Ho$_5$Sn$_3$, the possibility of spin frustration cannot be ruled out~\cite{palumbo}. This is also supported by the fact that $\theta_p$  is small and positive despite the occurrence of AFM transition at around 28 K. To address the issue, we  performed magnetic relaxation measurement at three temperatures, $T_i$ = 8 K, 20 K and 32 K.  The sample was cooled  from 150 K under $H$ = 100 Oe. On reaching a particular $T_i$, the field was withdrawn, and $M$ was measured as a function of time ($t$). At $T = $ 32 K, which is well above $T_N$, very weak time dependence of $M$ was observed.  However, as $T_i$ is lowered, the relaxation becomes stronger and as large as 14\% change in $M$ is observed in 60 min at $T_i$ = 8 K. The relaxation data are plotted in Fig.~\ref{relx} with  the time axis in the logarithmic scale. A linear curve indicates a logarithmic variation of $M$ with time~\cite{viscosity,gaunt} as given by the following equation,
\begin{equation}
\centering
 M(t) = M(0) - \mathcal{R}\ln (1 + t/\tau)~.
 \label{relax}
 \end{equation}
Here $\mathcal{R}$ is called the magnetic viscosity, $M(0)$ is the magnetization at $t$ = 0, and $\tau$ is some characteristics time scale. The value of $\mathcal{R}$ is found to be 0.76 emu/mol for the 8 K data, while it is 0.04 emu/mol at 20 K~\cite{Guy1978}. Logarithmic variation of $M$ with $t$ is not unusual in magnetic systems, and it is traditionally ascribed to the thermal activation of domain or domain walls over free energy barriers~\cite{gaunt}. The logarithmic dynamics originates from the distribution of energy barriers. 
%%%%%%%%%%%%%%%%%%%%%%%%%%%%%%%%%%%%%%%%%%%%%%%%%%%%%%%%%%%%%%%%%%%%%%%%%%%%%%%%%%%%
%%%%%%%%%%%%%%%%%%%%%%%%%%%%%%%%%%%%%%%%%%%%%%%%%%%%%%%%%%%%%%%%%%%%%%%%%%%%%%%%%%%%%%%
\par
We have also measured the field-cooled-field-stop memory on the sample to identify the possible glassy magnetic phase in the system~\cite{GdCu}. In this measurement, the samples is cooled under a field with intermediate stops. After reaching the lowest temperature, the sample is heated back again, and the system having a glassy magnetic phase shows an anomaly at the stopping points. However, our measurement does not record any such anomaly at the stopping points within the accuracy of the magnetometer. Therefore, a possible spin-freezing may be ruled out for Ho$_5$Sn$_3$.

\section{Discussions}
\label{discussions}
This combined theoretical and experimental work on Ho$_5$Sn$_3$ brings out several new and exciting  facts. first, the system is found to be susceptible to the applied magnetic field. We identify a clear spin flop transition, where the critical field follows a power-law dependence on temperature. The $M(T)$ curve flattens at higher values of $H$. Interestingly, we find an almost $T$ independent $M$ below 10 K, when measured under 50 kOe of field. This indicates that the system attains a  magnetic state at higher fields ($> H_C$)  markedly different from the low field state. The compound's ground state's magnetic structure is quite complex with three propagation vectors (two incommensurate sine modulated and one collinear commensurate).  The metamagnetism possibly arises from the development of FM component in the sine modulated structure and/or by the spin-flop transition of the commensurate structure. The metamagnetic transition shows field hysteresis indicating first order nature of the transition.   
\par
The DFT calculation clearly indicate the competition between FM and AFM states. The energy difference between the FM and AFM states are small ($\sim$ 37 meV), when the moments are aligned along the $c$ direction. It is to be noted that powder neutron diffraction data~\cite{ho5sn3} tell us that moment direction is indeed along the $c$ axis of the crystal. We also see that the ground state's nature depends upon the on-site Coulomb interaction, , $U_{\rm eff}$. This further strengthen the fact that AFM and FM states are energetically quite close.

\par
We find a low value of $\theta_p$, which is much lower than $T_N$. A difference between magnetic ordering temperature and $\theta_p$ can occur due to competing magnetic interactions, and few notable examples among intermetallic compounds are, Lu-doped Gd$_5$Ge$_4$~\cite{mudryk}, Pr$_5$Ge$_3$ and Nd$_5$Ge$_3$~\cite{nirmala}, Pr$_2$NiSi$_3$~\cite{pakhira1} Ho$_2$Ni$_{0.95}$Si$_{2.95}$~\cite{pakhira2} etc. A vanishing $\theta_p$ due to competing interaction is reported for Ca(VO)$_2$(PO$_4$)$_2$ and Sr(VO)$_2$(PO$_4$)$_2$~\cite{nath}, where the  magnetic frustration arises from the FM intra-chain and AFM inter-chain couplings. Similarly, in case of Ho$_2$Ni$_{0.95}$Si$_{2.95}$, the competing magnetic interaction arises due to the Ho-Ho nearest neighbour and next-nearest neighbour interactions.   Ho$_5$Sn$_3$ has a hexagonal crystal structure, and there remains the possibility that the intra-layer (within $a-b$ plane) and inter-layer (between $a-b$ planes) magnetic interactions compete with each other.

\par
An important finding of the present study is the slow dynamics of  $M$ at low temperature. Several of systems with coexisting magnetic phases show large relaxation due to the lack of thermodynamical equilibrium. The common examples are Gd$_5$Ge$_4$~\cite{gd5ge4-sbroy}, doped CeFe$_2$~\cite{cefe2-sbroy}, Pr$_{0.67}$Ca$_{0.33}$MnO$_3$~\cite{prcamno3}, where phase separation occurs due to the first-order phase transition. In addition, large relaxation is also observed in various metallic spin glasses~\cite{spin-glass}. Interestingly, Ho$_5$Sn$_3$  neither shows a phase coexistence due to first-order transition nor it has a  spin frozen state~\cite{pnas}. The compound also shows large irreversible susceptibility as evident from the bifurcation of ZFC and FC magnetizations.

\par
Notably, the FM intermetallic compounds TbPdIn and DyPdIn show  slow dynamics~\cite{RPdIn} akin to Ho$_5$Sn$_3$. The metastability in TbPdIn and DyPdIn was found to be associated with the effect of domain-wall pinning. We find that the sample shows a significant coercive field at 2 K ($\sim$ = 2.6 kOe), which is an indication of the presence of mageto-ctystalline anisotropy. The anisotropy is also supported by our  DFT based  calculations. We argue that the slow dynamics and the irreversible susceptibility in Ho$_5$Sn$_3$ is likely to be associated with the domain wall pinning. It is a well-known fact that magneto-crystalline anisotropy assists the pinning of domain walls~\cite{pinning,domainwall}. Our first principle calculation indicates that the nature of the magnetic ground-state depends critically on $U_{eff}$ (see Table~\ref{T1}). Therefore, a small change in $U_{eff}$ due to defect, disorder or inhomogeneity can alter the magnetic state locally over the spatial dimension of the sample. Such local fluctuations in the presence of anisotropy can pin the domain walls resulting slow relaxation.   
 
\par
Ho$_5$Sn$_3$ shows sizable MCE in terms of $\Delta S$, which can be attributed to the large moment of Ho and the associated metamagnetism. The peak in the value of $\Delta S$ occurs at around 30 K, and the $\Delta S$ vs $T$ curve shows half maximum values at 15 and 45 K. The sample does not show any field hysteresis in this temperature range, and it can be useful for low-$T$ refrigeration with minimum energy loss due to field cycling. The $M$ vs $T$ data also do not show any temperature hysteresis, and it is favorable for a magnetocaloric material.  Most importantly, we observe a change in sign of   $\Delta S$, around 8 K, which is evident from both the $\Delta S (T)$ and  $\Delta S(H)$ plots.  The $\Delta S$ is found to be positive  at 3 K and 6 K, while it is negative at all other temperatures above. An important feature of the $\Delta S(H)$ data is that irrespective of its positive or negative values, clear signature of metamagnetism at $H_C$ is present as long as $T < T_N$. The positive  $\Delta S$ (see 3 K and 6 K data)  is weakly field dependent (particularly above about 20 kOe), while negative   $\Delta S$ (mainly 15 and 27 K data) show strong enhancement in magnitude with increasing field. This is also clear from the $\Delta S (T)$ data, where hardly any change in the value of  $\Delta S$ is visible in the positive regime ($T <$ 7 K).

\par
The sample also shows large RC and RCP values, and they are comparable to the values reported for many other RE intermetallic compounds and alloy showing MCE at low temperature. Table~\ref{mce-comp} compared the values of entropy change of few other RE systems showing large MCE at low temperature. It is clearly seen that the RCP of Ho$_5$Sn$_3$ is reasonably good among the other compositions.  The large values of RC and RCP is due to the broad peak of the  $\Delta S (T)$ plot around $T_N$. The sample can be useful for  magnetic refrigeration around 30 K, which is an important temperature for cryogenic applications as the boiling point for liquid hydrogen is very close to this temperature. Ho$_5$Sn$_3$ shows both conventional ($\Delta S <$ 0)  and inverse ($\Delta S <$ 0) MCE, which can be technologically advantageous to build efficient refrigerator. For a conventional MCE material, the demagnetization process is only utilized for cooling. However, in the presence of both conventional and inverse MCE, magnetization (increasing $H$) and demagnatization (decreasing $H$) cycles can be used for magnetic cooling by suitably designing the refrigeration cycle~\cite{zhang1,zhang2,landazabal}. This can effectively increase the cooling efficiency of the material.

\par
In conclusion, both theoretical and experimental studies indicate a competition between FM and AFM states in the presently studied Ho$_5$Sn$_3$ compound. The sample shows a significant magnetocaloric effect which comprises both conventional and inverse change in magnetic entropy under a magnetic field. Such coexistence of conventional and inverse effect can be quite advantageous in building magnetic refrigerator using Ho$_5$Sn$_3$ as active material, particularly in the cryogenic temperature range.

\bibliography{smbiblio}

\end{document}